\documentclass[aps, prb, preprint, floatfix, amsmath]{revtex4-1}

\usepackage{natbib}
\usepackage{graphicx}
\usepackage{physics}
\usepackage{braket}
\usepackage{float}
\usepackage{mhchem}
\usepackage{hyperref}

\begin{document}
\title{Radiative lifetimes of dipolar excitons in double quantum-wells}

\author{Yotam Mazuz-Harpaz}
\author{Kobi Cohen}
\author{Boris Laikhtman}
\author{Ronen Rapaport}
\affiliation{Racah Institute of Physics, The Hebrew University of Jerusalem, Jerusalem 9190401, Israel.}
\author{Ken West}
\author{Loren N. Pfeiffer}
\affiliation{Department of Electrical Engineering, Princeton University, Princeton, New Jersey 08544, USA.}

\date{\today}

\begin{abstract}

Spatially indirect excitons in semiconducting double quantum wells have been shown to exhibit rich collective many-body behavior that result from the nature of the extended dipole-dipole interactions between particles. For many spectroscopic studies of the emission from a system of such indirect excitons, it is crucial to separate the single particle properties of the excitons from the many-body effects arising from their mutual interactions. In particular, knowledge of the relation between the emission energy of indirect excitons and their radiative lifetime could be highly beneficial for control, manipulation, and analysis of such systems. Here we study a simple analytic approximate relation between the radiative lifetime of indirect excitons and their emission energy. We show, both numerically and experimentally, the validity and the limits of this approximate relation. This relation between the emission energy and the lifetime of indirect excitons can be used to tune and determine their lifetime and their resulting dynamics without the need of directly measuring it, and as a tool for design of indirect exciton based devices.

\end{abstract}

\maketitle

\section{Introduction}

\begin{figure}
\centering
\includegraphics[width=0.45\textwidth]{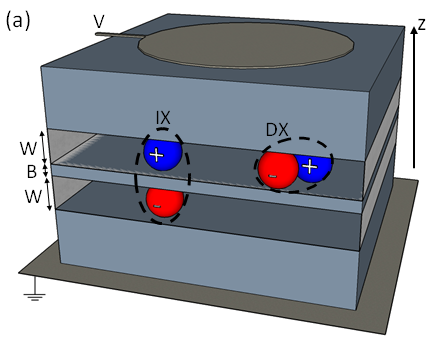}
\includegraphics[width=0.45\textwidth]{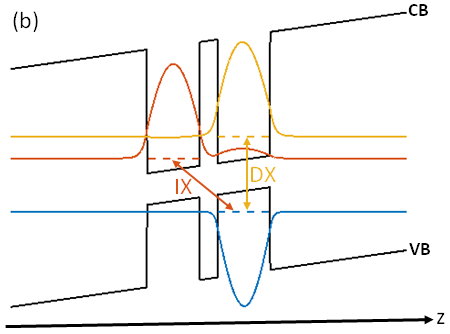}
\caption{
\label{fig:Illustration} 
(a) A simplified illustration of the typical realization of IXs in semiconductor DQW heterostructure. Two quantum wells of width $W$ are separated by a thin central barrier layer marked as $B$.  A metallic gate is positioned on the top of the sample, allowing the induction of electric field in the $z$ direction. Bound pairs of $e$ (red sphere) and $h$ (blue sphere) are created in the wells by optical excitation and can occupy the DX state - where both particles reside in the same well, or in the IX state - where each of them resides in a different well. (b) A typical energy diagram, illustrating the conduction (CB) and valence (VB) energy bands along the $z$ axis, the wavefunctions of the $h$ (blue line) and the $e$ occupying the DX (yellow line) and IX (red line) states, and their corresponding energies (dashed lines). The external electric field is responsible to the tilt of the energy bands, making the IX state lower in energy than the DX state.}
\end{figure}

An indirect exciton (IX) is a coulomb-bound complex of an electron ($e$) and a hole ($h$) where the opposite charges of the bound complex are spatially separated in two parallel layers with a tunneling barrier between them, as is depicted in Fig. \ref{fig:Illustration}. Such IXs are usually formed by optical excitation of an electrically biased double quantum well (DQW) heterostructure  \cite{rapaport_experimental_2007,butov_condensation_2004,butov_cold_2007}. This spatial $e$-$h$ separation leads to a long radiative lifetime and to a large electrical dipole moment of the IX. Due to the combination of these two properties, IXs give a unique opportunity to observe and study interesting cold interacting low-dimensional fluids. In recent years, an intensive experimental and theoretical effort focused on IX fluids in GaAs based DQWs revealed many intriguing many-body phenomena, such as spontaneous pattern formation \cite{butov_macroscopically_2002,snoke_long-range_2002,rapaport_charge_2004,butov_formation_2004}, spin textures \cite{high_spin_2013}, interaction-induced particle correlations \cite{laikhtman_exciton_2009,shilo_particle_2013}, molecular IX complexes \cite{schinner_confinement_2013,cohen_vertically_2016}, as well as evidence for complex collective phases \cite{butov_towards_2002,combescot_bose-einstein_2007,shilo_particle_2013,alloing_evidence_2014,stern_exciton_2014,cohen_dark_2016}. On the other hand, recent progress in the techniques for control and manipulation of IXs led to demonstration of various kinds of complex device functionalities such as trapping schemes \cite{rapaport_electrostatic_2005,hammack_trapping_2006,kowalik-seidl_tunable_2012,schinner_confinement_2013,alloing_optically_2013,schinner_confinement_2013}, flow control, IX transport and routing \cite{high_control_2008, violante_dynamics_2014,winbow_electrostatic_2011,cohen_remote_2011,lazic_scalable_2014,violante_dynamics_2014}, and spin transport \cite{leonard_spin_2009,kowalik-seidl_long_2010}.

Due to these recent advancements, cold IX fluids started attracting the interest of a wider scientific community, and new experiments on dipolar fluids of IXs in newly emerging systems have been performed recently. These include bilayer two-dimensional transition metal DiChalcagonide systems \cite{rivera_observation_2015}, bilayer graphene \cite{li_negative_2016,li_excitonic_2016} and polaritonic systems \cite{cristofolini_coupling_2012,rosenberg_electrically_2016} among others. 

To fully understand and control the various properties of such optically generated IX fluids, it is very important to have a good insight on their intrinsic dynamics.  A key property of the IX dynamics is the radiative lifetime. This radiative lifetime due to the $e$-$h$ optical recombination is the dominant loss process in such systems, and can in principle be measured directly by time-resolved measurement of the decay of their photoluminescence after a pulse excitation \cite{phillips_theoretical_1989,golub_long-lived_1990,alexandrou_electric-field_1990,charbonneau_transformation_1988}. However, as many experiments are done in a steady state under continuous-wave excitation and do not involve a direct lifetime measurement, a method of inferring the lifetime from other measurable quantities can be very useful. In particular, when designing a sample, a complex device, or an experiment, a prior knowledge and understanding of such recombination dynamics could be essential. Previous theoretical works have proposed various general approaches allowing numerical calculation of the IX's lifetime \cite{galbraith_exciton_1989,kamizato_excitons_1989,lee_excitonic_1989,dignam_exciton_1991,linnerud_exciton_1994,takahashi_effect_1994,soubusta_excitonic_1999,szymanska_excitonic_2003,arapan_exciton_2005,sivalertporn_direct_2012}. In our previous work \cite{shilo_particle_2013} we have presented a simple approximated analytic model according to which the radiative lifetime of an IX is simply related to its emission energy. This model allows inferring relative radiative lifetimes of IXs within a limited range of experimental conditions, from their corresponding eigen-energies, without the need for complex numerical calculations. These eigen-enegies could in turn be easily measured - for example, from the emission spectra - or be calculated numerically using rather simple computer solvers. 

In the model of Ref. \onlinecite{shilo_particle_2013}  the recombination lifetime of an IX is inversely proportional to the squared overlap integral between the envelop wavefunctions of the $e$ and the $h$:
\begin{equation}\label{eq:idlifetime}
	\tau_{id}=\frac{\tau_d}{\left| \braket{\psi_e|\psi_h} \right|^2}
\end{equation}
where $\tau_{id}$, $\tau_{d}$ are the lifetimes of the IX and of the direct exciton (DX), respectively and $\psi_e$, $\psi_h$ are the envelop wavefunctions of the $e$ and the $h$ along the DQW growth ($z$) direction, respectively. Fig. ~\ref{fig:Illustration} schematically illustrates both the DX and IX states in a DQW under an applied electric field along the $z$-axis.

Under a weak enough external electric field, the lowest energy $e$ and $h$ envelope wavefunctions can be approximated as a linear combination of the corresponding lowest energy flat-band SQW wavefunctions \cite{bastard_wave_1988}:
\begin{align}
	\psi^e(z) &= c^e_l\psi^e_l(z) + c^e_r\psi^e_r(z) \\
	\psi^h(z) &= c^h_l\psi^h_l(z) + c^h_r\psi^h_r(z).
\end{align}
Initially, increasing the field will only effect the coefficients and not the validity of the approximation itself. However, when the field becomes strong enough such that the potential drop across each QW is of the order of the difference between the ground energy in the SQW and its first excited state energy, this approximation breaks. Such strong field leads to a substantial admixture of excited states, shifting the peak of the wavefunction away from the center of the corresponding QW. Another approximation can be made in cases where the effective mass of the $h$ is significantly larger than the effective mass of the $e$, and under a large enough external field. Under such circumstances, the lowest energy $h$'s wavefunction can be well approximated by only one SQW wavefunction, i.e., $\psi^h(z)= \psi^h_r(z)$ in the case of an electric field applied in the positive $z$-direction, as is illustrated in Fig. \ref{fig:Illustration}b. As a result, if the external field is strong enough, the computation of the overlap integral $\braket{\psi_e|\psi_h}$ of Eq. \ref{eq:idlifetime} can be approximated by integration of the $e$'s wavefunction only inside the right QW in which the $h$'s wavefunction is strongly confined. Thus, there is an intermediate range of  electric field values in which both of the above approximations should hold to a good accuracy.  Within this range, the computation of the recombination lifetime is reduced to the computation of $c^e_r$ - the amplitude of the $e$'s wavefunction in the $h$'s well:
\begin{equation}
	\tau_{id}=\frac{\tau_d}{|c_r^e|^2}
\end{equation}

As was shown in Ref.~\onlinecite{shilo_particle_2013}, diagonalizing the Hamiltonian for the electron in the basis of $\psi_l^e$ and $\psi_r^e$, yields:
\begin{equation} \label{eq:cr}
	c_r^e = \frac{t}{E_d-E_{id}}
\end{equation}
where $E_d$ and $E_{id}$ are the energies of the DX and IX respectively and $t$ is the following tunnelling matrix element:
\begin{equation} \label{eq:trnasmatel}
	t=\braket{\psi_l^e|(T+V_e(z))|\psi_r^e}-\braket{\psi_l^e|\psi_r^e}E_0
\end{equation}
with $T$ the kinetic energy, $V_e(z)$ is the electron's DQW potential, and $E_0$ the ground state energy of a non-interacting electron in a SQW potential. The expression in Eq. ~\ref{eq:cr} was obtained under the assumption that all other contributions to the potential energy (i.e., the external electric field, the interaction of the electron with neighboring IXs and its coulomb interaction with the $h$) contribute only negligible corrections to $t$. Additionally, it assumes that the tunneling matrix element $t$ is small compared to the energy difference between the DX and the IX, i.e. 
\begin{equation}
	t\ll \frac{1}{2}(E_d-E_{id}).
\end{equation}
Under the limitations and assumptions mentioned above, the radiative lifetime of an IX can approximately be expressed as \cite{shilo_particle_2013}:
\begin{equation} \label{eq:Lifetime}
	\frac{1}{\tau_{id}}=\frac{1}{\tau_d}\frac{t^2}{\left(E_d-E_{id}\right)^2}.
\end{equation}

In this work we numerically test the accuracy and validity limits of Eq. ~\ref{eq:Lifetime} and provide an experimental confirmation that it holds to a good accuracy in the expected validity range. 

The structure of this paper is as follows: in Sec. ~\ref{sec:numerics} we compare that analytic result to a numerical calculation using a coupled Schr\"odinger-Poisson solver under a mean-field approximation of the interactions between IXs. In Sec. ~\ref{sec:experiment} we present experimental results confirming the validity of Eq. ~\ref{eq:Lifetime} and of the underlying approximations. In Sec. ~\ref{seq:conclusion} we summarize our results and their conclusions.

		\section{Comparison to numerical calculations} \label{sec:numerics}

To check the validity, accuracy, and applicability limits of Eq. ~\ref{eq:Lifetime}, we first compared its predictions with numerical calculations of the overlap between the envelop wavefunctions of the $e$ and the $h$, using a one-dimensional, self-consistent, Schr\"odinger-Poisson solver \cite{harrison_quantum_2000}. In the numerical model, we assumed a mean-field approximation for the IX-IX interactions, where in-plane dipolar correlations\cite{laikhtman_exciton_2009} were neglected, as well as the binding interaction between the $e$ and the $h$. For any given applied field $F$ and IX density $n$, we obtained from the solver the $e$ and $h$ envelope wavefunctions and numerically computed their overlap integral. The IX radiative lifetime is inversely proportional to the square of this overlap integral, and thus it can be calculated up to a multiplicative constant. More importantly, the ratio between the lifetime of the DX and the IX can also be calculated in this way as:
\begin{equation}
	\frac{\tau_{d}}{\tau_{id}} = \frac{\left| \braket{\psi_e^{(1)}|\psi_h^{(1)}} \right|^2}{\left| \braket{\psi_e^{(2)}|\psi_h^{(1)}} \right|^2}
\end{equation}
where the superscripts mark the corresponding quantum number of the $e$ and $h$ energy levels in the DQW. Substituting into Eq. \ref{eq:Lifetime} we get the following equation:
\begin{equation} \label{eq:numError}
	\frac{\left| \braket{\psi_e^{(1)}|\psi_h^{(1)}} \right|^2}{\left| \braket{\psi_e^{(1)}|\psi_h^{(2)}} \right|^2} = \frac{t^2}{\left(E_d-E_{id}\right)^2}.
\end{equation}
Since the same numerical solver also yields $E_d-E_{id}$, the accuracy of this equality can be used to check the accuracy of Eq. ~\ref{eq:Lifetime}, after plugging in the transition matrix element $t$. 

The transition matrix element $t$ can be approximated numerically using the SQW wavefunctions $\psi_r$ and $\psi_l$, according to Eq. ~\ref{eq:trnasmatel}. We carried this calculation for a semi-infinite narrow DQW structure having a $4nm$-wide central barrier, and for different realistic GaAs DQW structures, all having $\ce{Al_{0.5}Ga_{0.5}As}$ barriers and $4nm$-wide central barriers. The GaAs QWs have widths of $8nm$, $10nm$, $12nm$ and $14nm$, respectively. The values of $t$ for these four GaAs DQWs are $0.36meV$, $0.2meV$, $0.13meV$, and $0.09meV$ respectively.

\begin{figure}
\centering
\includegraphics[width=0.45\textwidth]{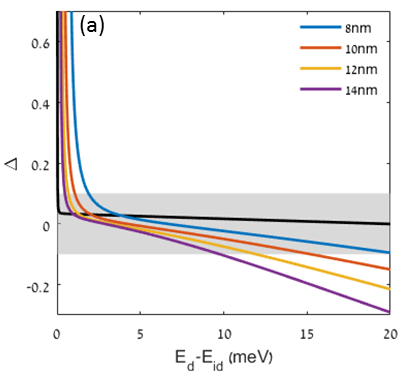}
\includegraphics[width=0.45\textwidth]{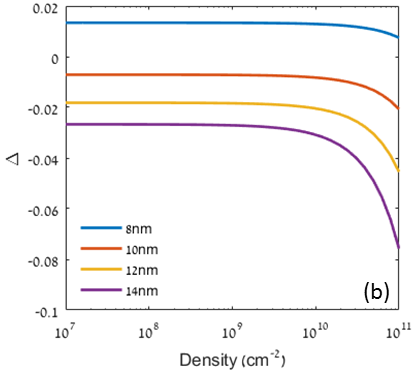}
\caption{\label{fig:Numerics} 
	(a) The relative error $\Delta$ of Eq. ~\ref{eq:numError}, defined in Eq. ~\ref{eq:error}, plotted against $E_d-E_{id}$ for four different widths (different W's) of a GaAs/$\ce{Al_{0.5}Ga_{0.5}As}$ DQW. The shaded area marks the region where $|\Delta|<0.1$ and the black curve is $\Delta$ calculated for the limit of a quasi-infinite, narrow DQW structure, also having a $4$nm-wide central barrier.
    (b) $\Delta$ as a function of the IX density $n$ while the external field is adjusted to keep the constant value of $E_d-E_{id}=5$ meV, for each of the four DQW structures.}
\end{figure}

The relative error $\Delta$ of Eq. ~\ref{eq:numError} is defined as $(r.h.s-l.h.s)/l.h.s$, i.e.
\begin{equation} \label{eq:error}
	\Delta \equiv \frac{t^2}{\left(E_d-E_{id}\right)^2}
    \times 
    \frac{\left| \braket{\psi_e^{(1)}|\psi_h^{(2)}} \right|^2}{\left| \braket{\psi_e^{(1)}|\psi_h^{(1)}} \right|^2} -1.
\end{equation}
This error is computed and presented in Fig. ~\ref{fig:Numerics}a as a function of $E_d-E_{id}$ for four different DQW structures differing by their well widths, for the single-IX case (i.e. where the IX-IX interaction is set to zero). Here the values of $E_d-E_{id}$ are determined solely by $F$. The dependence of the relative error $\Delta$ on $E_d-E_{id}$ is in agreement with the expected limits of validity mentioned in the previous section: where $E_d-E_{id}$ is very small, the penetration of the $h$'s wavefunction into the $e$'s QW is significant and Eq. \ref{eq:Lifetime} over-estimates the radiative lifetime. Once the field is increased such that $E_d-E_{id}$ becomes larger than $t$, the error drops rapidly and stays low for quite a wide range of $E_d-E_{id}$ values. At even larger values of $E_d-E_{id}$, both the $e$'s and the $h$'s wavefunctions distort in opposite directions and their overlap integral is further diminished. The analytic model does not take this distortion into account (as each of the basis wavefunctions used $\psi_l$ and $\psi_r$ is of a flat bottom QW), and so it underestimates the lifetime, resulting in a relative error which grows with $E_d-E_{id}$ in the negative direction. As expected, this distortion increases with the width of the QWs, leading to a wider $E_d-E_{id}$ range having a high relative accuracy of Eq. \ref{eq:Lifetime} for narrower DQWs.

We note that this calculation neglect the $e-h$ coulomb attraction, which tends to reduce the above distortion of the single particle wavefunctions with increasing $F$. In this sense the error values presented in in Fig. ~\ref{fig:Numerics}a are an overestimate of the expected error of Eq. \ref{eq:Lifetime}.

Fig. ~\ref{fig:Numerics}b presents the relative error as a function of the IX density $n$, for a fixed value $E_d-E_{id}=5$ meV. This is done by setting different values of $n$ and finding the corresponding values of $F$ to keep $E_d-E_{id}$ constant (this method, named 'constant energy line method', was extensively used in our previous experimental works \cite{shilo_particle_2013,cohen_dark_2016}). 
As seen in the figure, the relatively low error values are maintained up to a high IX density of $n\simeq10^{11}$cm$^{-2}$. Above this value, the inhomogeneous distribution of the IX charge density along the z-axis of the QWs is large enough to induce large deviations of the $e$ and $h$ wavefunctions from the single particle wavefunctions, leading to a decrease in the accuracy of Eq. ~\ref{eq:Lifetime}.

These numerical calculations demonstrate that Eq. \ref{eq:Lifetime} is a good approximation over a significant range of applied electric fields and IX densities and can be tested in experiments, as we show in the next section.

	\section{Comparison to experiments} \label{sec:experiment}
    
\subsection{Experimental details}

\begin{figure*}
\centering
\includegraphics[width=1\textwidth]{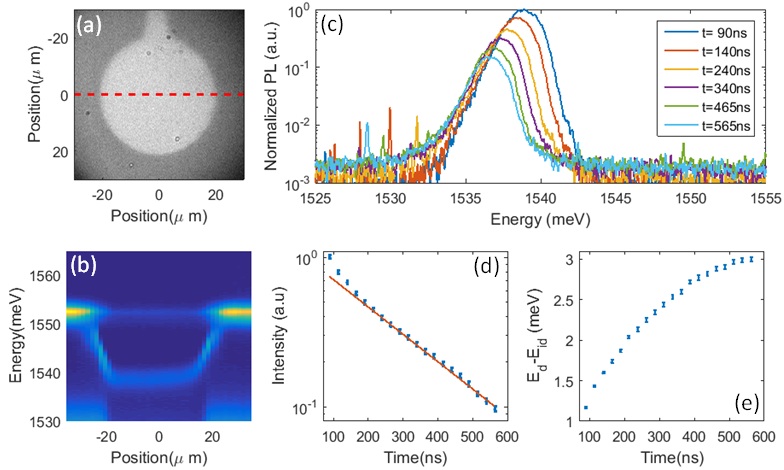}
\caption{\label{fig:AllRawData} 
	(a) A real-space image of the circular electrostatic trap used in this work. 	
    (b) A spectral-spatial image of the emission from the sample as a function of the position of a CW laser excitation along the dashed red line drawn in (a), with a reversed voltage set to $1V$, an excitation power of $1500$nW and temperature of $1.8K$. The trapping potential is flat around the center of the trap with $E_d-E_{id}\simeq 12$meV.
    (c) Exemplary time resolved and spatially integrated spectra at different times after the laser excitation. (d) Integrated intensity $I$ and (e)  $(E_d-E_{id})$ as a function of time after the laser pulse extracted from the data in (c). All time frames are integrated over a $25$ns-wide window.}
\end{figure*}

The experimental scheme is rather similar to that presented in few of our previous works \cite{shilo_particle_2013,cohen_dark_2016}. The sample, positioned in an optical $\ce{^{4}He}$ cryostat, is a $12/4/12$nm
$GaAs/Al_{0.5}Ga_{0.5}As/GaAs$ DQW structure grown on an $n^+$-doped GaAs substrate and with a $10$nm-thick, semi-transparent Ti electrode positioned on its top \cite{cohen_dark_2016}. The overall thickness of the sample is about $3\mu m$. An electric field between the top electrode and the doped substrate creates the energy band tilt required for the formation of IXs and allows their trapping as presented in fig ~\ref{fig:AllRawData}a-b. The sample temperature is maintained at $T=1.8$K.

A population of IXs is being excited using a non-resonant pulsed laser having wavelength of $775nm$ and pulse duration of $300ps$, under a fixed applied electric bias. We then study the decay of the population after the pulse excitation, by time-resolved measurement of the light emitted by recombination of the optically active (i.e 'bright') IXs, using a fast-gated ICCD camera (Princeton Instruments PIMAX). 

\subsection{Results}

Exemplary time resolved, spatially integrated spectra at different times after the laser excitation are presented in Fig. ~\ref{fig:AllRawData}c, for a fixed applied bias of 1V. The spectral lines have a tail to the long wavelength side, similarly to previous results. Fig. ~\ref{fig:AllRawData}d,e present the integrated intensity and the IX energy as a function of time after the excitation. The observed redshift during the decay results from the decrease of the interaction energy as the IX density decreases \cite{shilo_particle_2013}. 

The radiative recombination of IXs can be described by a simple rate equation:
\begin{equation}
	I(t)\propto\frac{\partial n}{\partial t} = \frac{n(t)}{\tau_{id}(t)}
\end{equation}
where $I$ is the measured emission intensity. Using $n(t)\propto\int_t^\infty I(t')dt'$, and expressing  $\tau_{id}(t)$ using Eq. \ref{eq:Lifetime} the following relation is obtained:
\begin{equation} \label{eq:main}
	I(t) \propto \frac{\int_t^\infty I(t')dt'}{\left( E_d-E_{id}(t) \right)^2} \equiv G(t)
\end{equation}
and both $I(t)$ and $G(t)$ can be directly expressed for every $t$ from the experimental results in Fig. ~\ref{fig:AllRawData}d,e. The assumptions made in this last derivation are the accuracy of Eq. ~\ref{eq:Lifetime} and that the recombination process is radiatively dominated.
Thus, if these assumptions are valid we expect that Eq. \ref{eq:main} should hold for our experimental results, i.e., we expect to find that $G(t)\propto I(t)$ for every $t$ along the whole decay. This is therefore a direct experimental test for Eq. ~\ref{eq:Lifetime}.  

\begin{figure}
\centering
\includegraphics[width=0.45\textwidth]{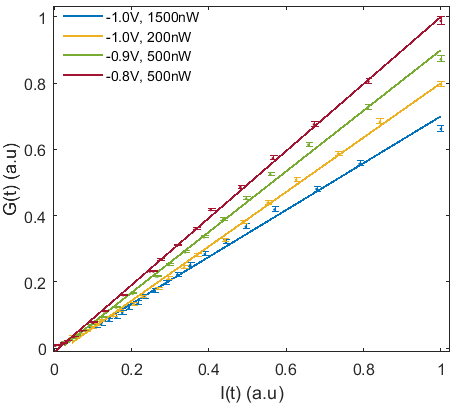}
\caption{\label{fig:DecayConst} Experimentally extracted $G(t)$ versus $I(t)$ for four different decay traces that were measured under fixed lattice temperature of $1.8K$ ($\pm0.1K$) and different external applied biases and laser powers. The solid lines are linear fits to the data. A clear linear dependence is observed, confirming the validity of Eq. ~\ref{eq:Lifetime} for this range of experimental parameters (The temporal dynamics in each of these measurements is from top-right to bottom-left).}
\end{figure}

Fig. ~\ref{fig:DecayConst} presents the experimentally extracted $G(t)$ versus $I(t)$ for four different decay traces that were measured under fixed lattice temperature of $1.8K$ ($\pm0.1K$) and different external applied biases and laser powers. The solid lines are linear fits to the data. A clear linear dependence is observed, confirming the validity of the above assumptions and thus the accuracy of Eq. ~\ref{eq:Lifetime} for this range of experimental parameters.

		\section{Conclusion} \label{seq:conclusion}
In this work we presented a numerical and experimental test of an approximated derivation relating the IX radiative lifetime to its energy. We showed that this relation holds well for a wide range of accessible experimental parameters and thus also confirmed our assumptions used in our previous works \cite{shilo_particle_2013,cohen_dark_2016}. We conclude that this relation can be very useful for future works studying IXs in similar bilayer systems, simplifying the analysis of dynamics of IX systems. In many such experiments, the lifetime is a key property whose assessment is not a straight-forward task under the required experimental settings. This difficulty could be relieved by the proposed method, which in principle requires a single calibration measurement to set right the proportion coefficient of Eq. ~\ref{eq:Lifetime} for the sample of interest.

As demonstrated above by the numerical simulation, Eq. ~\ref{eq:Lifetime} is only valid and accurate enough in an intermediate range of IX energies: the basic picture described by the theoretical model becomes valid only when the separation between the energies of the indirect and the direct excitons is large enough. As the separation is increase further, the accuracy of the model gradually deteriorate and the error of Eq. ~\ref{eq:Lifetime} grows. Between these two ends, lies the range of IX energies where the error is relatively low and Eq. ~\ref{eq:Lifetime} can be used as a good approximation.

We believe that this method can be easily modified for other IX structures that are currently being explored, such as IXs in bilayers of other material systems.

\begin{acknowledgments}
We would like to thank Masha Vladimirova for fruitful discussions. We would also like to acknowledge financial support from the German DFG (grant No. SA-598/9), from the German Israeli Foundation (GIF I-1277-303.10/2014), and from the Israeli Science Foundation (grant No. 1319/12). The work at Princeton University was funded by the Gordon and Betty Moore Foundation through the EPiQS initiative Grant GBMF4420, and by the National Science Foundation MRSEC Grant DMR-1420541.

\end{acknowledgments}

\bibliography{Zotero}
\bibliographystyle{ieeetr}

\end{document}